\documentclass[useAMS,usenatbib]{mn2e}
\usepackage{graphicx}
\usepackage{times}
\usepackage{amssymb}
\usepackage{natbib}

\def\gsim { \lower .75ex \hbox{$\sim$} \llap{\raise .27ex \hbox{$>$}} }
\def\lsim { \lower .75ex \hbox{$\sim$} \llap{\raise .27ex \hbox{$<$}} }

\begin{document}

\title{Orbital Eccentricity as a probe of Thick Disk Formation Scenarios}

\author[Sales et al.]{
\parbox[t]{\textwidth}{
Laura V. Sales$^{1}$,
Amina Helmi$^{1}$,
Mario G. Abadi$^{2}$,
Chris B. Brook$^{3}$,
Facundo A. G\'omez$^{1}$,
Rok Ro{\v s}kar$^{4}$,
Victor P. Debattista$^{3}$,
Elisa House$^{3}$,
Mathias Steinmetz$^{5}$,
\'Alvaro Villalobos$^{1}$
}
\\
\\
$^{1}$ Kapteyn Astronomical Institute, P.O. Box 800, Groningen, The Netherlands.\\
$^{2}$ Universidad Nacional de C\'ordoba, Laprida 854, 5000 C\'ordoba, Argentina 
and Instituto de Astronom\'{\i}a Te\'orica y Experimental, Conicet, Argentina.\\
$^{3}$ Centre for Astrophysics, University of Central Lancashire, Preston, PR1 2HE, UK.\\
$^{4}$ Department of Astronomy, University of Washington, Box 351580, Seattle, WA 98195.\\
$^{5}$ Astrophysikalisches Institut Potsdam, An der Sternwarte 16, Potsdam 14482, Germany.\\
}

\maketitle

\begin{abstract}
  We study the orbital properties of stars in four (published)
  simulations of thick disks formed by: $i)$ accretion from disrupted
  satellites, $ii)$ heating of a pre-existing thin disk by a minor
  merger, $iii)$ radial migration and $iv)$ gas rich mergers. We find
  that the distribution of orbital eccentricities are predicted to be
  different for each model: a prominent peak at low eccentricity is
  expected for the heating, migration and gas-rich merging scenarios,
  while the eccentricity distribution is broader and shifted towards
  higher values for the accretion model. These differences can be
  traced back to whether the bulk of the stars in each case is formed
  {\it in-situ} or is {\it accreted}, and are robust to the
  peculiarities of each model. A simple test based on the eccentricity
  distribution of nearby thick disk stars may thus help elucidate the
  dominant formation mechanism of the Galactic thick disk.
\end{abstract}

\section{Introduction}
\label{sec:intro} 

Several mechanisms have been proposed to explain the formation of
thick disks in galaxies \citep[see][]{majewski93}. However it is still
unclear by which of these mechanisms thick disks preferentially
form. This is despite the fact that more than 25 years have passed
since it was first detected in the Milky Way \citep{gilmore_reid83},
and that it has been established that this component appears to be
ubiquitous in late-type systems \citep[e.g.][and references
therein]{yoachim08}.

Amongst the scenarios proposed to explain the formation of the thick
disks are the direct accretion of stars from disrupted satellites
\citep[e.g. ][]{abadi03b}, the thickening of a pre-existing thin disk
through a minor merger
\citep[e.g. ][]{quinn93,villalobos08,kazantzidis08}, the scattering or
migration of stars by spiral arms
\citep[e.g.][]{schoenrich09a,roskar08a,schoenrich09b}, and
{\it in-situ} trigered star formation during/after gas-rich mergers
\citep[e.g.][]{brook05,bournaud07}.

Even though studies of external galaxies have been fundamental to
establish the statistical properties of thick disks, it is likely that
only for the Galactic thick disk we will be able to unravel its
evolutionary path. For example, measurements of the phase-space
coordinates for nearby thick disk stars allows reconstruction of their
orbits, which contain imprints of the dynamical history, while their
chemical abundances encode information about their sites of
origin. Time is ripe to delve into more detailed predictions for the
above-mentioned scenarios, because these have reached a level of
maturity and detail that they warrant and permit a nearly direct
comparison to observations.

In this {\it Letter} we investigate how the orbits of thick disk stars
can be used to distinguish between the various formation channels.  In
particular, we focus on the predicted eccentricity distributions.  We
expect our findings to be applied soon to samples of nearby thick
disk stars from SEGUE \citep{yanny09,smith09} and RAVE (Steinmetz et
al. 2006, Breddels et al.\ submitted), and in the long term, to the
{\it Gaia} dataset which will provide much more accurate information
for much larger samples of stars spanning a wide range of distances
from the Sun. The orbital eccentricity-test should help to elucidate
the dominant mechanism by which the Galactic thick disk
formed.\nocite{steinmetz06}

\vspace*{-0.5cm}

\begin{table*}
\begin{center}
  \caption{Relevant parameters for each of the simulations. (1) model,
    (2) Virial mass, (3) Bulge/Spheroid mass (4) Disk mass
    (thin+thick), (5) Thin disk scale-height (from a double power law
    fit to the vertical mass profile), (6) Thick disk scale-height,
    (7) Radial scale length of the thin disk, (8) softening lengths,
    (9) reference to the original articles. For comparison, estimated
    values for the Milky Way have also been included. For the
    cosmological simulations, the virial masses are defined as the
    mass enclosed within the radius where the local density falls
    below $\Delta=100$ times the critical density of the universe.  A
    Hubble constant $H_0=70\, \rm Mpc^{-1} km/s$ is assumed when
    necessary. Disks scale-heights have been determined by fitting
    double-exponential laws to the vertical stellar density profile
    in the cylindrical shell $2<R/R_d<3$.}
\begin{tabular}{|c|c|c|c|c|c|c|c|c|}
\hline
\hline
Scenario & $M_{vir}$ & $M_{bulge}$ & $M_{disk}$ & $z_0$ (thin) &  $z_0$ (thick) & $R_d$ & $\epsilon$ & Reference\\
 & $[10^{10} \rm M_\odot]$ & $[10^{10} \rm M_\odot]$ & $[10^{10} \rm M_\odot]$ & $[\rm kpc] $ & $[\rm kpc]$ & $[\rm kpc]$ & $[\rm kpc]$ &\\
\hline
{\it accretion} & 87 & 6.7 & 2.8 & 0.5 & 2.3 & 4.1  & 0.50 & Abadi et al. (2003a,b)\\
{\it heating} & 50 & -- & 1.2  & -- & 1.2 & -- & 0.01  & Villalobos \& Helmi (2008)\\
{\it migration} & 100 & 4.8 & 3.0 & 0.3 & 0.9 & 3.5 & 0.05 & Roskar et al. (2008) \\
{\it merger} & 71 & 2.1 & 3.4  & 0.3 &  1.0 & 2.9 & 0.40  & Brook et al. (2004)  \\
{Milky Way} & 60-200 & 1. & 7-10 & 0.3 & 0.9 & 3.5 & -- &\citet{turon08} \\
\hline
\end{tabular}
\label{table:simu}
\end{center}
\end{table*}

\section{Numerical Experiments}
\label{sec:simu}

We have gathered four existing numerical simulations of
late-type galaxies that, having all developed a thick
disk component, clearly differ in the dominant formation
mechanism. These are:
\begin{enumerate}
\item accretion and disruption of satellites \citep{abadi03b}, 
\item disk heating by a minor merger \citep{villalobos08},
\item radial migration via resonant scattering \citep{roskar08a},
\item {\it in-situ} formation during/after a gas-rich merger
\citep{brook04,brook05}.
\end{enumerate}

\subsection{Thick disk formation models}
\label{sec:description}

Because each of the simulations mentioned above have been already
introduced in the literature, here we will only review their main
relevant features, referring the reader to the original papers for
further details.  Table \ref{table:simu} summarizes their key
parameters. 

\subsubsection{Accretion scenario}
\label{ssec:accretion}

\citet{abadi03b} showed that within the $\Lambda$CDM paradigm, the
accretion of stars from disrupting satellites in approximately
co-planar orbits may give rise to an old thick disk component that
comprises about one-third the mass of the much younger thin disk.

In our sample we include the galaxy presented in \citet{abadi03b},
which formed in a cosmological N-body/SPH simulation.  This object,
with a virial mass of the order of that of the Milky Way, was
selected from a low-resolution simulation of a large volume of the
Universe; and later re-simulated with much higher resolution. In this
high resolution run, the mass per baryonic particle is $\sim 3 \times
10^6 M_\odot$. The final mass for the thick disk in this galaxy
(derived via a dynamical decomposition) is $1.1 \times 10^{10}
M_\odot$.

\vspace{-0.25cm}
\subsubsection{Heating scenario}
\label{ssec:heating}

In this model, a thick disk is formed by the dynamical heating that is
induced by a massive satellite merging with a primordial, rotationally
supported thin disk. This scenario has been explored recently by
e.g. \citet{villalobos08, kazantzidis08}, who have shown that 5:1 mergers
and with a wide range of orbital inclinations generate thick disks whose
properties are in reasonable agreement with observations. In such a
model, the bulk of stars that end up in the thick disk originate from
the primordial disk rather than from the accreted satellite
\citep{villalobos08}.

In our analysis we include one of the numerical experiments presented
in Villalobos \& Helmi (2008). In these simulations, the mass ratio
between the satellite and the host is 0.2 and its initial orbit is
prograde and inclined by 30$^o$ with respect to the host disk. The
mass per stellar particle in the simulation is $m_p = 1.2\times 10^5
M_\odot$, and the thick disk has a final mass of 1.2$\times
10^{10} M_\odot$. It is important to clarify that only a small fraction of
the thin disk component is present at the end of the simulation ($\sim 15-20\%$
the mass of the original disk). This
implies that, for this remnant to be the thick disk of a late-type
galaxy, a new thin disk should form later from the cooling of fresh
gas. This will lead to structural changes in the thicker component,
which are not considered here. Nevertheless, if the growth of the new
disk is adiabatic, then many characteristics, and in particular the
eccentricities, are not expected to be dramatically different
\citep{villalobos-thesis}.

\subsubsection{Radial Migration scenario}
\label{ssec:migration}

Stars in the thin disk may be trapped onto resonant corotation with
spiral arms, and may migrate inwards and outwards along the spiral
waves approximately conserving their angular momenta (and hence
eccentricity) and without leading to significant heating in the disk
\citep{selwood_binney02}.  However, since the vertical velocity
dispersion of stellar disks correlates with their surface brightness
\citep{kregel05}, the {\it radial migration} of stars from the inner
regions (kinematically hotter) will result in the formation of a
thicker disk component.

Although this process has not been formally proposed as a thick disk
formation scenario, numerical simulations suggest that a modest thick
component may be built.  Therefore, we include in our sample the
simulation presented in \citet{roskar08a} ran with the goal of
characterizing the migration that takes place in galactic disks. The
simulation starts with a dark matter halo of $\sim 10^{12} M_\odot$
where 10\% of this mass is in the form of a hot halo gas
component. This gas is allowed to cool and form stars
self-consistently, mimicking the quiescent growth of disk galaxies,
over a period of 10~Gyr.  The initial mass resolution is $10^5
M_\odot$ for the baryons, and stellar particles have on average masses
$3 \times 10^4 M_\odot$.

\subsubsection{Gas-rich merger scenario}
\label{ssec:merger}

The last scenario we explore consists in the formation of a thick
rotating component during an active epoch of gas rich mergers in the
past history of a galaxy \citep{brook04,brook05}. This formation
channel differs fundamentally from the {\it accretion} model because
the bulk of thick disk stars are born {\it in-situ} rather than being
accreted from satellites.  In this sense, this scenario might show
certain similarities with the {\it heating} model.  However, the
latter requires the existence of a thin disk at early times, $z \geq
1$, in contrast to the {\it merger} scenario where the stars are
already born in a hotter component.

Here we analyze the simulated galaxy introduced in \citet{brook04}.
It formed in a semi-cosmological N-body/SPH simulation that includes
heating/cooling of gas, star formation, feedback and chemical
enrichment. Its dark halo has a quiescent merger history after $z
\!\sim\!2$, and a final baryonic content of the galaxy is $\sim 5
\times 10^{10} M_\odot$ (see Table~\ref{table:simu}). The mass per
baryonic particle is $\sim 2 \times 10^{5} M_\odot$. The mass of the
thick disk in this galaxy is $\sim 2.2 \times 10^9 M_\odot$,
identified as old stars ($8.5<\rm age<10.5$) with relatively high
rotation velocity ($V_{\rm rot}>50$ km/s).

\begin{center}
\begin{figure}
\includegraphics[width=84mm]{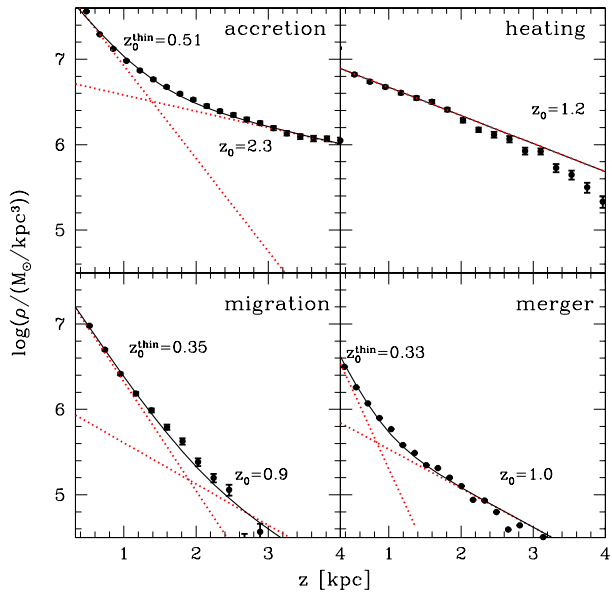}
\caption{Vertical density profile of stars for each of the scenarios
   discussed in Section \ref{sec:description}.  The best fit
   mass-weighted double-exponential profiles are shown with black solid lines, and
   the individual contributions of the ''thin'' and ''thick''
   components are indicated by the red dotted curves. Notice that
   there is no significant thin disk in the {\it heating} scenario,
   thus only the vertical profile for the thick component is
   present. For the accretion scenario kinematical cuts
   have been applied in order to avoid contamination from the stellar
   halo (see text for details)}.
\label{fig:zprof}
\end{figure}
\end{center}

The scenarios described in this Section are capable of producing a
rotationally supported hot component whose properties resemble the
`thick disks' in galaxies. However, the relative preponderance of such
thick components does vary from galaxy to galaxy in our
simulations. This is illustrated in Figure \ref{fig:zprof}, where we
show with solid dots the vertical mass profiles for each case in a
cylindrical shell $2<R/R_d<3$; which minimizes the contribution from
bars and bulges. Additionally, in the accretion model, particles
identified as ``spheroid'' in Abadi et al. (2003b) have been removed
(see Section 3). The error bars correspond to the rms obtained
from one hundred bootstrap re-samples of the data, and they are
generally smaller than the dot's sizes. The black solid lines show
the best-fit double exponential profile found for each galaxy,
together with its decomposition into the ``thin'' and the ``thick''
disk contributions, in red dotted lines.  This decomposition is
generally robust, although correlations exist between the relative
density of each component and the scale-height of the thick disk. The
scale-heights $z_0$ obtained by minimizing $\chi^2$ are quoted in each
panel, and the typical errors are of order $\sim 5-10\%$ for
$z_0^{thin}$ and $\sim 20-30\%$ for the thick component. For the
migration scenario, the uncertainties are larger due to the stronger
dominance of the thin disk, and are 15\% and 60\% for $z_0^{thin}$ and
$z_0$, respectively. Nevertheless, in all cases, the results presented
below are robust to changes in the value of $z_0$ within the
uncertainties.

Differences in the relevance of the thick component
depend not only on the net efficiency of the respective formation
process, but may also be influenced by the different initial
conditions and simulation techniques (e.g.\ only the {\it accretion}
and {\it merger} scenarios actually account for the full cosmological
framework).  Because of these basic differences between simulations,
global properties such as the mass, rotation and size of each formed
thick disk are expected to be diverse. Our focus, however, is on
contrasting the specific dynamical properties of the stars in the
thick component for each case, and in particular, their orbital
eccentricities, which as we shall see below, are fundamentally related
to the physical mechanism by which this component was built.

In order to facilitate comparisons between the thick disks in our
galaxies and in particular, also to that of the Milky Way, we re-scale
the radial and vertical distances of the stellar particles in each
galaxy by their corresponding thin disk scale-lengths and thick disk
scale-heights\footnote{For the simulation by Villalobos \& Helmi
  (2008) we assume the scale-length to be that of the Milky Way thin
  disk: $R_d=3.5$~kpc, although our conclusions do not fundamentally
  depend on this choice.}. In what follows, we will focus our
attention on ``solar neighbourhood volumes'', equivalent to
cylindrical shells between two and three scale-radii of the thin disk
($2<R/R_d<3$). For comparison, $R_\odot/R_d \sim 2.2-2.4$, assuming a
scale radius of $R_d = 3.5$ kpc for the Milky Way.

\subsection{Modelling of the orbits}

Kinematical surveys such as RAVE, SEGUE and ultimately {\it Gaia}
provide phase-space coordinates of stars around the position of
the Sun. This instantaneous information may be used to recover their
plausible past orbits. This requires modelling the (unknown) Galactic
potential, and possibly its evolution, which implies that the orbital
parameters derived for each star generally suffer from a certain
degree of uncertainty, even if measurement errors are neglected.

On the other hand, our numerical simulations allow us to track in time
each particle, with full orbits that are known. Nevertheless
we prefer to mimic observations, and therefore we use the present-day
position and velocity of each stellar particle as initial conditions
for the integration of their orbits in the best-fit potential of
their host galaxy.  We model each galaxy as a four component system
with an NFW \citep{nfw97} dark halo, a Hernquist profile 
\citep{hernquist90} for the bulge and two Miyamoto-Nagai disks 
\citep{miyamoto_nagai} corresponding to the thin and thick
disks contributions. The mass associated with each of these components
is known for each simulation (see Table~\ref{table:simu}), and their
various scale-lengths are chosen by requiring a good match to the
circular velocity profile of the system up to a distance of
20~kpc\footnote{For the migration and merger scenarios only the {\it
total} (thin + thick) mass of the disk is known. For these cases the
relative mass ratio between the thin and thick components is also a
free parameter.}.  This is along the lines of previous work, where the
circular velocity of the Milky Way is often used to constrain the
model parameters \citep[e.g.][]{helmi06}.

\begin{center}
\begin{figure}
\includegraphics[width=84mm]{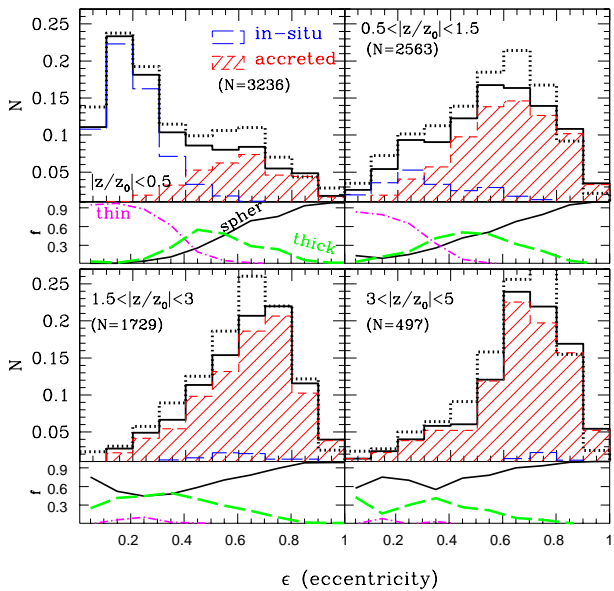}
\caption{Eccentricity distribution of all stellar particles in a
  cylindrical shell with $2<R/R_d<3$ for the {\it accretion} scenario
  \citep{abadi03b}. The various panels are for different heights
  above/below the plane (normalized to the thick disk scale-height,
  $z_0$).  The thick solid black line shows the total distributions
  per $z$-bin, while blue-empty and red-shaded histograms distinguish
  between {\it in-situ} and accreted stars. The effect introduced by
  the numerical integration of the orbits is rather small: the
  $\epsilon$-distribution obtained from direct tracking of the
  particles orbits in the simulation is shown in dotted black
  line. The fractional contributions from the thin disk (magenta
  dot-dashed), thick disk (green long-dashed) and the spheroid (solid
  black) to each eccentricity bin are shown below each histogram. The
  number of particles $N$ included in each box is also quoted.}
\label{fig:abadi_allz}
\end{figure}
\end{center}

We define eccentricity as $\epsilon=(r_{ap}-r_{pe})/(r_{ap}+r_{pe})$,
where $r_{ap}$ and $r_{pe}$ correspond, respectively, to the apo- and
pericenter distance of the last orbit of each particle. With this
definition, for a circular velocity curve modeled with $\sim 10\%$
accuracy, the eccentricities obtained by numerical integration show a
scatter $\pm 0.1-0.2$ around their true value with no systematic
trends (see Fig.~\ref{fig:abadi_allz}). Larger deviations are found as
the eccentricity increases.

\vspace{-0.3cm}
\section{Results}
\label{sec:analysis}

Baryons in galaxies are generally sorted in several components: a
bulge, a disk (thin+thick) and a more extended and diffuse spheroidal
distribution, the stellar halo, that might extend well beyond the
luminous edge of the disk.  Although each of these components has
defining characteristics, some of its properties may change smoothly
from component to component.  For example the eccentricity of the
orbits changes as we move away from the disk plane into the realms of
the stellar halo. This can be seen in Figure \ref{fig:abadi_allz},
where the eccentricity distribution of all stellar particles within a
cylindrical radii $2<R/R_d<3$ is plotted for different heights
above/below the plane. Vertical distances are normalized to the
scale-height of the thick disk, $z_0$.  The eccentricity distributions
for stars formed {\it in-situ} (defined as those born within a
distance of 20 kpc from the main progenitor) and for those {\it accreted}
are given by empty long-dashed blue and shaded red histograms
respectively.  For comparison, the eccentricity distribution measured
directly from the simulation (i.e. by tracking individual particle
orbits) is given by the dotted histogram. This shows that no
systematic errors are introduced by the orbital integration in the
model host potential.

The lowermost bin $|z/z_0|<0.5$ is largely dominated by thin disk
stars with circular motions, as can be seen by the strong peak around
$\epsilon \sim 0.15$ in the top left panel of Figure
\ref{fig:abadi_allz}.  These stellar particles formed {\it in-situ}
through local conversion of gas settled in a disk, into stars (blue
empty histogram) within the main galaxy.  As we move away from the
plane the thick disk gains importance and the eccentricity
distributions become flatter as the fraction of accreted stars
increases. Further above and below the plane, the distribution is
dominated by particles from the spheroidal component, with even higher
characteristic eccentricities, i.e $\epsilon \sim 0.7$. These changes
in the relative preponderance of the thin, thick and spheroidal
components can be seen as their fractional contribution (magenta
dot-dashed, green long-dashed and black solid, respectively) to each
eccentricity on the bottom panel of all $|z|$-bins.  Here, we have
used the dynamical decomposition performed in \citet{abadi03b} to
assign stars to a given component.

This galaxy, introduced in Abadi et al. (2003a,b), has a large stellar
spheroid that contains more than $\sim$70\% of the total luminous
mass. Figure \ref{fig:abadi_allz} shows that its contribution
dominates the high eccentricity bins at all heights above/below the
plane. To highlight the properties of the thick disk, and also to
avoid confusion on the interpretation of our results and its
comparison with other simulated galaxies (lacking such a prominent
spheroidal component), we will exclude in what follows any stellar
particle that have been assigned to the spheroid by the analysis
performed in Abadi et al. (2003b).\nocite{abadi03a,abadi03b}

\begin{center}
\begin{figure}
\includegraphics[width=84mm]{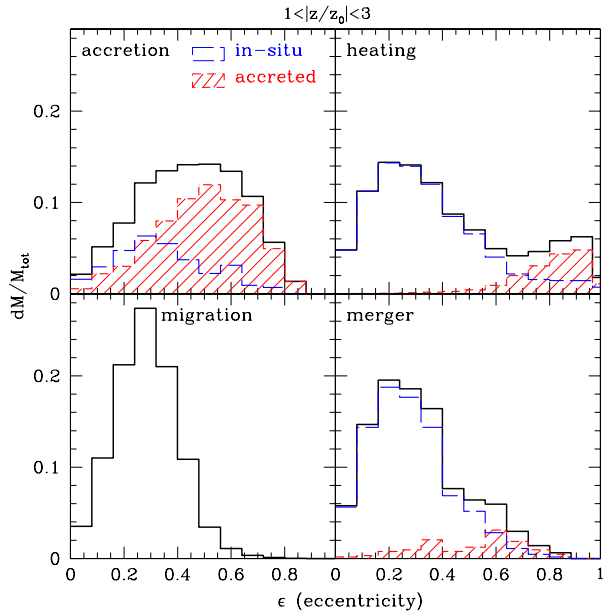}
\caption{
Comparison of the eccentricity distributions of each thick disk
formation model for stars in the range 1-3 (thick-disk) scale-heights
and cylindrical distance $2 < R/R_d < 3$. 
The color and line coding are the same as introduced in
  Figure \ref{fig:abadi_allz}.}
\label{fig:e_all}
\end{figure}
\end{center}

Figure \ref{fig:e_all} shows how the eccentricity distributions vary
according to the different formation channels of the thick disk.  Each
panel corresponds to one particular model: {\it accretion} (top left),
{\it heating} (top right), {\it migration} (bottom left) and {\it
merger} (bottom right). When relevant, the contributions from stars
formed ``{\it in-situ}'' or ``{\it accreted}'' have been highlighted.
In order to minimize the contribution from thin disk stars, we have
focused on the vertical bin $1<|z/z_0|<3$.  On the other hand, to
avoid contamination from the spheroids in our simulations (this is
unlikely to be important for the Galactic stellar halo because of its
very low density), we only consider stars with rotational velocity
$v_\phi > 50$~km~s$^{-1}$.  This corresponds to average azimuthal
velocity at which there is a clear excess of stars with velocities
larger than this threshold compared to the distribution at $v_\phi <
-50$~km~s$^{-1}$ in our simulations. Under the assumption of a
non-rotating spheroidal component, this criterion will minimize the
contribution of the stellar halo in our samples. Nonetheless, we have
checked that our results are not strongly sensitive to this
assumption.  The $v_\phi > 50$~km~s$^{-1}$ cut removes 33 per cent of
the stars in Brook's model, but has a negligible effect in the
simulation by Villalobos \& Helmi and Ro{\v s}kar et al.  due to the
suppressed contribution of satellite accretion. Recall that for the
Abadi's galaxy we have removed all the spheroid identified by the
dynamical decomposition.  Cuts on the cylindrical radii ($2<R/R_d<3$)
also help to elude the contributions from central bars and bulges.

Figure \ref{fig:e_all} shows that the eccentricity distributions of
stellar particles in these ``solar neighbourhood'' regions, and between 
one and three thick disk scale-heights above/below the plane, are different
according to each model.  For the {\it accretion}
scenario, the distribution is very broad, with a median eccentricity $
\langle \epsilon \rangle \sim 0.5$ \citep[in good agreement with the
accreted component in][]{reed08}. On the other hand, the heating of a
pre-existing thin disk by a minor merger gives rise to a bimodal
distribution. The dominant peak is at low eccentricity $\epsilon \sim
0.2 - 0.3$ and associated to the stars from the progenitor disk, while
the second peak at $\epsilon \sim 0.8$, is brought by the disrupted
satellite.  {\it Radial migration} tends to preserve the initial (low)
eccentricity distribution, with only one peak present at $ \langle
\epsilon \rangle\sim 0.2$ and with a sharp cut-off at $\epsilon \sim
0.6$. Finally, in the {\it merger} scenario a prominent peak around
$\epsilon \sim 0.2$ is visible which is associated to stars formed
{\it in-situ} during the epoch of active gas-rich mergers. But like
for the {\it accretion} scenario, accreted stars from infalling
satellites contribute to a high-eccentricity tail.

Interestingly, Figure \ref{fig:e_all} shows, as expected, that stars
formed {\it in-situ} have low eccentricity orbits regardless the
mechanism ({\it heating, migration, merger}) that places them at their
current height above/below the plane. On the other hand, accreted stars from
satellites always dominate the high eccentricity tails of the
distributions. Changes in the relative proportion of {\it in-situ}
versus accreted stars drive the differences seen in the histograms of
each thick disk formation scenario.  In other words, the analysis of
the stellar eccentricities off the plane helps to unravel whether
the bulk of stars was formed locally in the main progenitor ({\it
  heating, migration, merger}) or, on the contrary, was accreted from
infalling satellites ({\it accretion}).

Kolmogorov-Smirnov tests performed over randomly generated
subsamples of stars selected from each simulation show that $\sim 150$
stars are enough to distinguish at 90\% confidence level between all
the scenarios investigated here. Moreover, in samples containing more
than $\sim 550$ stars, the probability that the stars are drawn from
the same distribution is found to be lower than 1\%. Clearly a smaller
number of stars ($\sim 200$) is sufficient to distinguish the
accretion model from the rest. Note, however, that these estimates are
only indicative since they are derived for particular realizations of
a general class of models.

It is important to recall that each of these simulations has produced
a galaxy with a different morphology. Furthermore the simulations
representative of the {\it heating} scenario as well as that of {\it
migration} have, by construction, a suppressed contribution from
accreted satellite galaxies. Nevertheless, the general behaviour of
{\it in-situ} vs {\it accreted} populations are robust to the
different simulations idiosyncrasies and can be traced back to the
different physical mechanisms related to where and how the stars were
formed. Therefore, we do not expect the global properties of the
eccentricity distribution (e.g. bimodality, high eccentricity tails
associated to accreted populations) to differ significantly, but only
in the details, in other realizations of the same models. Although we
have focused on ``solar-neighbourhood'' regions ($2<R/R_d<3$), our
conclusions do not depend fundamentally on this choice.

\section{Conclusions}
\label{sec:concl}

In this {\it Letter} we have analyzed four numerical simulations of
galaxies hosting a thick disk component of fundamentally different
origin: $(i)$ accretion of satellites, $(ii)$ heating of a
pre-existing disk by a 5:1 mass-ratio merger, $(iii)$ radial migration
by resonant scattering and $(iv)$ gas-rich mergers at high-redshift.
  
We have compared the eccentricity distributions predicted by these
different models for stellar particles in the ``solar neighbourhood'',
i.e. located in a cylindrical shell of radius $2<R/R_d<3$ and with
heights $1 <|z/z_0|<3$. Thick disk stars formed {\it in-situ} have low
orbital eccentricities $\epsilon \sim 0.2-0.3$, independently of the
mechanism that brought them high above/below the plane: gas-rich mergers,
heating or migration. On the other hand, and again regardless of the
particular model, accreted stars always dominate the high-eccentricity
tail of the distributions. Therefore, the characterization of the
eccentricity distribution of the thick disk can be used to establish
if this component was formed by the accretion of satellites (Abadi et
al. 2003) or alternatively locally within the main galaxy
\citep{brook05,villalobos08,roskar08a}. However, given the various
limitations of our set of simulations, we cannot claim that it will be
possible to make an unequivocal classification among models $(i)-(iv)$
based only on eccentricity.

Nevertheless, the differences between the orbital eccentricities of
{\it in-situ} and accreted stellar particles are encouraging in view
of the various kinematic surveys mapping our Galaxy today and in the
near-future. We believe that with a reasonable guess of the Milky Way
potential, the analysis of the eccentricity distribution of thick disk
stars at approximately 1-3 scale-heights should shed light
on the formation path of the Galactic thick disk.

\vspace*{-0.5cm}
\section*{Acknowledgements}
The authors thank the hospitality of the KITP, Santa Barbara, where
this work was started, and in particular Juna Kollmeier for
encouragement and for her contagious enthusiasm. LVS and AH gratefully
acknowledge NWO and NOVA for financial support.  This research was
supported in part by the National Science Foundation under Grant
No. PHY05-51164. We also thank the anonymous referee for useful
suggestions and comments.

\bibliography{references}

\begin{thebibliography}{}

\bibitem[\protect\citeauthoryear{{Abadi}, {Navarro}, {Steinmetz} \&
  {Eke}}{{Abadi} et~al.}{2003a}]{abadi03a}
{Abadi} M.~G.,  {Navarro} J.~F.,  {Steinmetz} M.,    {Eke} V.~R.,  2003a, \apj,
  591, 499

\bibitem[\protect\citeauthoryear{{Abadi}, {Navarro}, {Steinmetz} \&
  {Eke}}{{Abadi} et~al.}{2003b}]{abadi03b}
{Abadi} M.~G.,  {Navarro} J.~F.,  {Steinmetz} M.,    {Eke} V.~R.,  2003b, \apj,
  597, 21

\bibitem[\protect\citeauthoryear{{Bournaud}, {Elmegreen} \&
  {Elmegreen}}{{Bournaud} et~al.}{2007}]{bournaud07}
{Bournaud} F.,  {Elmegreen} B.~G.,    {Elmegreen} D.~M.,  2007, \apj, 670, 237

\bibitem[\protect\citeauthoryear{{Brook}, {Gibson}, {Martel} \&
  {Kawata}}{{Brook} et~al.}{2005}]{brook05}
{Brook} C.~B.,  {Gibson} B.~K.,  {Martel} H.,    {Kawata} D.,  2005, \apj, 630,
  298

\bibitem[\protect\citeauthoryear{{Brook}, {Kawata}, {Gibson} \&
  {Freeman}}{{Brook} et~al.}{2004}]{brook04}
{Brook} C.~B.,  {Kawata} D.,  {Gibson} B.~K.,    {Freeman} K.~C.,  2004, \apj,
  612, 894

\bibitem[\protect\citeauthoryear{{Gilmore} \& {Reid}}{{Gilmore} \&
  {Reid}}{1983}]{gilmore_reid83}
{Gilmore} G.,  {Reid} N.,  1983, \mnras, 202, 1025

\bibitem[\protect\citeauthoryear{{Helmi}, {Navarro}, {Nordstr{\"o}m},
  {Holmberg}, {Abadi} \& {Steinmetz}}{{Helmi} et~al.}{2006}]{helmi06}
{Helmi} A.,  {Navarro} J.~F.,  {Nordstr{\"o}m} B.,  {Holmberg} J.,  {Abadi}
  M.~G.,    {Steinmetz} M.,  2006, \mnras, 365, 1309

\bibitem[\protect\citeauthoryear{{Hernquist}}{{Hernquist}}{1990}]{hernquist90}
{Hernquist} L.,  1990, \apj, 356, 359

\bibitem[\protect\citeauthoryear{{Kazantzidis}, {Bullock}, {Zentner},
  {Kravtsov} \& {Moustakas}}{{Kazantzidis} et~al.}{2008}]{kazantzidis08}
{Kazantzidis} S.,  {Bullock} J.~S.,  {Zentner} A.~R.,  {Kravtsov} A.~V.,
  {Moustakas} L.~A.,  2008, \apj, 688, 254

\bibitem[\protect\citeauthoryear{{Kregel}, {van der Kruit} \&
  {Freeman}}{{Kregel} et~al.}{2005}]{kregel05}
{Kregel} M.,  {van der Kruit} P.~C.,    {Freeman} K.~C.,  2005, \mnras, 358,
  503

\bibitem[\protect\citeauthoryear{{Majewski}}{{Majewski}}{1993}]{majewski93}
{Majewski} S.~R.,  1993, \araa, 31, 575

\bibitem[\protect\citeauthoryear{{Miyamoto} \& {Nagai}}{{Miyamoto} \&
  {Nagai}}{1975}]{miyamoto_nagai}
{Miyamoto} M.,  {Nagai} R.,  1975, \pasj, 27, 533

\bibitem[\protect\citeauthoryear{{Navarro}, {Frenk} \& {White}}{{Navarro}
  et~al.}{1997}]{nfw97}
{Navarro} J.~F.,  {Frenk} C.~S.,    {White} S.~D.~M.,  1997, \apj, 490, 493

\bibitem[\protect\citeauthoryear{{Quinn}, {Hernquist} \& {Fullagar}}{{Quinn}
  et~al.}{1993}]{quinn93}
{Quinn} P.~J.,  {Hernquist} L.,    {Fullagar} D.~P.,  1993, \apj, 403, 74

\bibitem[\protect\citeauthoryear{{Read}, {Lake}, {Agertz} \&
  {Debattista}}{{Read} et~al.}{2008}]{reed08}
{Read} J.~I.,  {Lake} G.,  {Agertz} O.,    {Debattista} V.~P.,  2008, \mnras,
  389, 1041

\bibitem[\protect\citeauthoryear{{Ro{\v s}kar}, {Debattista}, {Stinson},
  {Quinn}, {Kaufmann} \& {Wadsley}}{{Ro{\v s}kar} et~al.}{2008}]{roskar08a}
{Ro{\v s}kar} R.,  {Debattista} V.~P.,  {Stinson} G.~S.,  {Quinn} T.~R.,
  {Kaufmann} T.,    {Wadsley} J.,  2008, \apjl, 675, L65

\bibitem[\protect\citeauthoryear{{Sch{\"o}nrich} \& {Binney}}{{Sch{\"o}nrich}
  \& {Binney}}{2009a}]{schoenrich09a}
{Sch{\"o}nrich} R.,  {Binney} J.,  2009a, \mnras, 396, 203

\bibitem[\protect\citeauthoryear{{Sch{\"o}nrich} \& {Binney}}{{Sch{\"o}nrich}
  \& {Binney}}{2009b}]{schoenrich09b}
{Sch{\"o}nrich} R.,  {Binney} J.,  2009b, ArXiv e-prints

\bibitem[\protect\citeauthoryear{{Sellwood} \& {Binney}}{{Sellwood} \&
  {Binney}}{2002}]{selwood_binney02}
{Sellwood} J.~A.,  {Binney} J.~J.,  2002, \mnras, 336, 785

\bibitem[\protect\citeauthoryear{{Smith}, {Evans}, {Belokurov}, {Hewett},
  {Bramich}, {Gilmore}, {Irwin}, {Vidrih} \& {Zucker}}{{Smith}
  et~al.}{2009}]{smith09}
{Smith} M.~C.,  {Evans} N.~W.,  {Belokurov} V.,  {Hewett} P.~C.,  {Bramich}
  D.~M.,  {Gilmore} G.,  {Irwin} M.~J.,  {Vidrih} S.,    {Zucker} D.~B.,  2009,
  ArXiv e-prints

\bibitem[\protect\citeauthoryear{{Steinmetz}, {Zwitter}, {Siebert}, {Watson},
  {Freeman}, {Munari}, {Campbell} \& {and 47 coauthors}}{{Steinmetz}
  et~al.}{2006}]{steinmetz06}
{Steinmetz} M.,  {Zwitter} T.,  {Siebert} A.,  {Watson} F.~G.,  {Freeman}
  K.~C.,  {Munari} U.,  {Campbell} R.,    {and 47 coauthors} 2006, \aj, 132,
  1645

\bibitem[\protect\citeauthoryear{{Turon}, {Primas}, {Binney}, {Chiappini},
  {Drew}, {Helmi}, {Robin} \& {Ryan}}{{Turon} et~al.}{2008}]{turon08}
{Turon} C.,  {Primas} F.,  {Binney} J.,  {Chiappini} C.,  {Drew} J.,  {Helmi}
  A.,  {Robin} A.~C.,    {Ryan} S.~G.,  2008, Technical report, {Galactic
  Populations, Chemistry and Dynamics}

\bibitem[\protect\citeauthoryear{{Villalobos}}{{Villalobos}}{2009}]{villalobos%
-thesis}
{Villalobos} {\'A}.,  2009, PhD thesis, University of Groningen

\bibitem[\protect\citeauthoryear{{Villalobos} \& {Helmi}}{{Villalobos} \&
  {Helmi}}{2008}]{villalobos08}
{Villalobos} {\'A}.,  {Helmi} A.,  2008, \mnras, 391, 1806

\bibitem[\protect\citeauthoryear{{Yanny}, {Rockosi}, {Newberg}, {Knapp},
  {Adelman-McCarthy}, {Alcorn}, {Allam} \& {and SEGUE team}}{{Yanny}
  et~al.}{2009}]{yanny09}
{Yanny} B.,  {Rockosi} C.,  {Newberg} H.~J.,  {Knapp} G.~R.,
  {Adelman-McCarthy} J.~K.,  {Alcorn} B.,  {Allam}   {and SEGUE team} 2009,
  \aj, 137, 4377

\bibitem[\protect\citeauthoryear{{Yoachim} \& {Dalcanton}}{{Yoachim} \&
  {Dalcanton}}{2008}]{yoachim08}
{Yoachim} P.,  {Dalcanton} J.~J.,  2008, \apj, 682, 1004

\end{thebibliography}

\end{document}